\documentclass[
reprint,
superscriptaddress,
amsmath,amssymb,
prl,
]{revtex4-2}
\usepackage{color}
\usepackage[T2A]{fontenc}			 
\usepackage[utf8]{inputenc}			 
\usepackage{csquotes}
\usepackage{tempora}
\usepackage[english]{babel} 
\usepackage{ulem}					 
\usepackage{graphicx}
\usepackage{dcolumn}
\usepackage{bm}
\usepackage[unicode=true,colorlinks=true,citecolor=blue,urlcolor=blue]{hyperref}

\newcommand{\ee}{\mathrm{e}}
\renewcommand{\i}{{\rm i}}
\renewcommand{\d}{\mathrm d}
\renewcommand{\Re}{\mathop{\rm Re}}

\usepackage{braket}

\begin{document}

\author{V.~O.~Kozlov}
\affiliation{Spin Optics Laboratory, St.\,Petersburg State University, 198504 St.\,Petersburg, Russia}
\affiliation{Laboratory for Crystal Photonics, St.\,Petersburg State University, 198504 St.\,Petersburg, Russia}

\author{N.~I.~Selivanov}
\affiliation{Photonics of crystals laboratory, St.\,Petersburg State University, 198504 St.\,Petersburg, Russia}

\author{C.~C.~Stoumpos}
\affiliation{Laboratory for Crystal Photonics, St.\,Petersburg State University, 198504 St.\,Petersburg, Russia}
\affiliation{Department of Metarials Science and Technology, University of Crete, 71003 Heraklion, Greece}

\author{G.~G.~Kozlov}
\affiliation{Spin Optics Laboratory, St.\,Petersburg State University, 198504 St.\,Petersburg, Russia}
\affiliation{Solid State Physics Department, St.\,Petersburg State University, Peterhof, 198504 St.\,Petersburg, Russia}
	
\author{V.~S.~Zapasskii}
\affiliation{Spin Optics Laboratory, St.\,Petersburg State University, 198504 St.\,Petersburg, Russia}
\affiliation{Laboratory for Crystal Photonics, St.\,Petersburg State University, 198504 St.\,Petersburg, Russia}

\author{Yu.~V.~Kapitonov}
\affiliation{Laboratory for Crystal Photonics, St.\,Petersburg State University, 198504 St.\,Petersburg, Russia}
\affiliation{Photonics Department, St.\,Petersburg State University, Peterhof, 198504 St.\,Petersburg, Russia}

\author{D.~S.~Smirnov}
	\affiliation{Ioffe Institute, 194021 St.\,Petersburg, Russia}
	\affiliation{Spin Optics Laboratory, St.\,Petersburg State University, 198504 St.\,Petersburg, Russia}
 
	\author{I.~I.~Ryzhov}
	\affiliation{Photonics Department, St.\,Petersburg State University, Peterhof, 198504 St.\,Petersburg, Russia}
	\affiliation{Spin Optics Laboratory, St.\,Petersburg State University, 198504 St.\,Petersburg, Russia}
	\affiliation{Laboratory for Crystal Photonics, St.\,Petersburg State University, 198504 St.\,Petersburg, Russia}
	
\begin{abstract}
We report on first observation of spin noise in a strongly birefringent semiconductor~-- halide perovskite single crystal MAPbI$_3$.
The observed spin noise resonance is ascribed to free holes with a record spin dephasing time of 4~ns. The spin dynamics is found to be  affected by the residual light absorption of the crystal providing renormalization of the Larmor frequency. Extended spin noise spectroscopy with rotating magnetic field allowed us not only to evaluate the $g$\nobreakdash-factor anisotropy, but also to distinguish two different spin subsystems tentatively associated to twinning of the crystal.
\end{abstract}

\title{Spin noise of a halide perovskite}

\maketitle
	
\section*{Introduction}

Spin noise spectroscopy (SNS) is an optical method of studying spin dynamics 
that rapidly expands, nowadays, the scope of its application~\cite{horn-sns-donor-bound-e-in-zno13,cronenberger-atomic-like-sn15,kamenskii-re-sns2020,fomin-spin-align-noise2020,sun-non-equil-sns-single-qw-telecom22,liu-sns-spin-one-system23}. 
Primarily, this experimental approach was {applied to atomic systems ~\cite{aleksandrov-mr-fr-noise-en81,crooker-sns-nature04}, that fully uncovered its uniqueness: the SNS probes the electron paramagnetic resonance in a non-perturbing manner away from the optical resonance of the studied medium.}
At the beginning of this century, the SNS was applied to semiconductors~\cite{oestreich-sns-gaas05,romer-sns-review07} and became popular as an efficient tool for studying magnetic resonance and spin dynamics of charge carriers in bulk~\cite{crooker-sn-gaas2009,horn-sns-donor-bound-e-in-zno13,cronenberger-atomic-like-sn15} and low-dimensional semiconductor structures~\cite{muller-mqw-sns08,crooker-sns-e-holes-qd2010,poltavtsev-sns-single-qw14,cronenberger-spatiotemporal-sns2019}.
{Since the sample is studied in the transparency region and most of the spin noise signal comes from the Rayleigh waist length of the beam~\cite{romer-spatially-resolved-sns09}, the SNS provides information about spin subsystem in a small volume determined by focusing parameters of the laser beam.} The SNS turned out to be useful also for resonant probing of optical transitions, when it could not be considered as nonperturbative~\cite{zapasskii-sns-review13,zapasskii-optical-sns13,glasenapp-sns-beyond-thermal-equil14,PhysRevB.90.085303,smirnov-optical-res-shift-sn-spectrosc20}.
Specifically, it usually demonstrates nontrivial dependence of the signal on the light power density~\cite{fomin-high-spin21,fomin-anomal-broad-sn-cs21,sun-non-equil-sns-single-qw-telecom22}, allows one to distinguish homogeneously and inhomogeneously broadened optical transitions~\cite{ma-osns-rb-gas-hom-inhom17,petrov-homogenization-doppler2018}, and provides a simple way for studies of the light-induced phenomena such as detection of  the ``optical'' magnetic field~\cite{ryzhov-sn-explores-local-fields16} or nuclear orientation~\cite{ryzhov-nuclear-dynamics-sns15,nuclear-spin-vladimirova2017}. In recent years, the SNS became applicable to the rare-earth-doped dielectric crystals~\cite{kamenskii-re-sns2020} and, more importantly, strongly birefringent materials~\cite{kozlov-sn-birefringent22}. The latter result may seem paradoxical, because the SNS implies detection of the Faraday rotation noise, whereas, in optically anisotropic media, the Faraday effect is strongly suppressed~\cite{shumitskaya-fr-phase-trans-perovskite-2023-preprint}. Still, in the semiconductor area, the application of SNS remained limited by non- or negligibly weakly birefringent crystals, mainly cubic GaAs and CdTe-based structures.

Recently, the {halide perovskite semiconductors}
attracted a lot of attention, mainly due to their outstanding properties for photovoltaics~\cite{kojima-organometal-perovskite-photovoltaic09,doi:10.1021/nl5048779,fu2019metal,zhao2022bilayer}. The deeper investigations revealed their peculiar spin properties, which may be promising for the spintronics~\cite{ping-spin-optotronic-perovkites18}. Specifically, spin-polarized exciton quantum beats~\cite{odenthal-spin-pol-q-beating-perovskites2017}, spin control of the lasing threshold~\cite{tang-perovskite-spin-laser22}, spin manipulation of charge carriers~\cite{belykh-coherent-spin-dynamics-e-h-perovskite19,kirstein-coherent-spin-dyn-e-2d-peapi-perovskites23} and nuclei~\cite{adma_Dortmund,kirstein2023evidencing} as well as  other related magneto-optical phenomena~\cite{shrivastava-polaron-spin-dyn-perovskite20,garcia-arellano-energy-tuning-e-spin-perovskites21,garcia-arellano-unexp-anisotropy-g-factor-perovskite22} were reported in the past few years. Typically, to study the spin properties of halide perovskites, methods based on optical excitation are used, such as pump-probe Faraday and Kerr-rotation~\cite{kirstein-spin-dyn-e-h-interact-nuclei-mapi-single-crystal22}, polarized photoluminescence~\cite{wang-spin-optoelectronic-devices-org-inorg-perovskites19}, and spin-flip Raman scattering~\cite{harkort-spin-flip-raman-e-h-2d-peapbi-perovskites23}.
{The inevitable absorption of the light in these methods drives the sample out from the thermodynamic equilibrium. The photoexcitation can lead to a change of the concentration or even of the type of charge carriers and modify their properties due to nonlinear effects.}
{The SNS is the method of choice for studying the unperturbed state of the spin subsystem of halide perovskites. A necessary condition for detecting the spin noise signal is the presence of a paramagnetic contribution to the Faraday rotation.} 
Our recent work~\cite{shumitskaya-fr-phase-trans-perovskite-2023-preprint} has shown that the MAPbI$_3$ (MA$^+$~=~CH$_3$NH$_3^+$) crystal exhibits the {paramagnetic (Curie-law-obeyed)} Faraday rotation, thus making it suitable for non-perturbative spin noise investigations. In this work, we make use of  applicability  of SNS to birefringent media to measure  spin noise in the MAPbI$_3$ crystal. Specific abilities of the SNS allowed us to observe the record long spin coherence time of around 4~ns, to measure anisotropy of the spin-carrier $g$\nobreakdash-factor, and to reveal spontaneous twinning of the crystal.

\section{Results}
\paragraph{Samples and characterization.}
	
The samples under study were MAPbI$_3$ single crystals grown by the counterdiffusion-in-gel method~\cite{selivanov-counterdiffusion-in-gel-growth-perovskite22}. The crystals from the same growth run were preliminary studied using the Faraday rotation (FR) method~\cite{shumitskaya-fr-phase-trans-perovskite-2023-preprint}. Temperature dependence of the detected FR signal unambiguously indicated its paramagnetic nature and, therefore, the presence of spin carriers potentially amenable to SNS. The spin noise signal was obtained at temperatures down to 3.5\,K using the standard SNS setup (see Methods and Supplementary Information (SI) for details of the setup and Refs.~\cite{zapasskii-sns-review13,glasenapp-polarimetric-sens13} for the general SNS specificity) including a tunable CW Ti:sapphire laser, closed-cycle cryostat, broadband balanced photoreceiver, and fast Fourier transform (FFT) based radiofrequency (RF) spectrum analyzer. {The measurements were carried out with the probe light energies 1.55--1.62~eV well below the bandgap of MAPbI$_3$. This ensured the observation of the Faraday rotation noise in the transmitted light and allowed us to make perturbation of the sample negligibly small.}

The MAPbI$_3$ crystal is known to be orthorhombic below 162~K~\cite{onoda-yamamuro-calorimetric-ir-phase-transition-perovskite90,kawamura-structural-study-perovskite02}. The decrease in crystal symmetry leads to appearance of birefringence, but does not affect opportunity of the spin-nose measurements~\cite{kozlov-sn-birefringent22}. To evaluate the balue of the birefringence, we performed special measurements: we have measured transmission spectrum of the sample sandwiched between crossed polarizers. At room temperature, the birefringence obtained from specytal period of the oscillating transmission was found to be $\Delta n \approx 0.02$ (see SI Fig.~1 for details).

\begin{figure}
		\centering
		\includegraphics[width=0.8\linewidth]{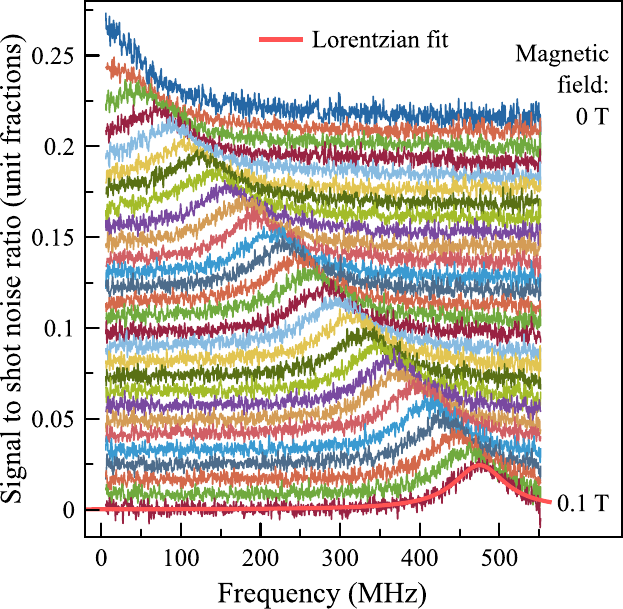}
		\caption{Magnetic field dependence of the spin noise spectrum.  Probe power 8~mW (2~mW transmitted), $\lambda = 769.2$~nm. The spectra are normalized to the shot noise of transmitted light, 
   see~\cite{poltavtsev-sns-single-qw14} for details.
		}
		\label{signal-vs-field} 
	\end{figure}

\paragraph{Spin noise signal: General features and magnetic-field dependence.}
{The spin noise was detected in transmitted light as a spectrum of fluctuations of the Faraday rotation in the range of radiofrequencies. The presence of spin carriers in the sample was manifested by appearance of a  peak at the Larmor frequency, with its position governed by the spin carrier $g$-factor and its homogeneous width controlled by dephasing time of the spin system.}
The spin noise spectrum of the MAPbI$_3$ at 3~K in external transverse magnetic field showed a single peak of Lorentzian shape (Fig.~\ref{signal-vs-field}) thus indicating smallness of its inhomogeneous broadening~\cite{crooker-sn-gaas2009,horn-sns-donor-bound-e-in-zno13,cronenberger-atomic-like-sn15}. The peak shifted linearly with magnetic field without significant shape variations up to 0.1~T. The linear fit of the peak position dependence on the magnetic field yielded the effective $g$\nobreakdash-factor of 0.35 (an example of the fit is shown in Fig.~\ref{signal-vs-field} for the largest magnetic field). The full width at half maximum of the Lorentzian peak, $\Delta\omega_L$$\approx$\,90~MHz, corresponded to the transverse spin relaxation time as long as $T_2 \approx 3.5$~ns. As the azimuth of the polarization plane was rotated, the peak changed its amplitude without changing its position or width, as expected for a birefringent crystal~\cite{kozlov-sn-birefringent22} (SI, Fig.~3b.).
	
	\paragraph{Dependence of spin noise on temperature, probe wavelength and power.}

With increasing temperature, the area of the peak decreased without changing its width (SI, Fig.~4), which {could indicate} a decrease in concentration of the {probed carriers. To some extent, this supposition is confirmed by measurements of photoinduced conductivity~\cite{pisoni-metallicity-conductivity-crossover-mapi-perovskite16}, which decreased with increasing temperature. Another possible reason is a decrease in the spin noise gain factor, specified by the ratio of inhomogeneous and homogeneous optical widths~\cite{kamenskii-re-sns2020}. However, this issue requires further investigation, which is beyond the scope of this paper.}
 
{The spin noise peak amplitude monotonically decreases with increasing probe light wavelength}. The signal becomes unobservable at wavelengths exceeding 800~nm (1.55~eV), and the total area of the peak rises up to 767 nm (1.62~eV). However, at shorter wavelengths, the transmitted power decreases dramatically due to the Urbach tail absorption, thus preventing further optical measurements. The peak width exhibits almost no variation while the probe wavelength is tuned from 800 to 767~nm (1.55 to 1.62 eV),
indicating practically no perturbation of the spin system by the probe light. {More details are presented in SI, Fig.~5.}

\begin{figure}
\centering
\includegraphics[width=\linewidth]{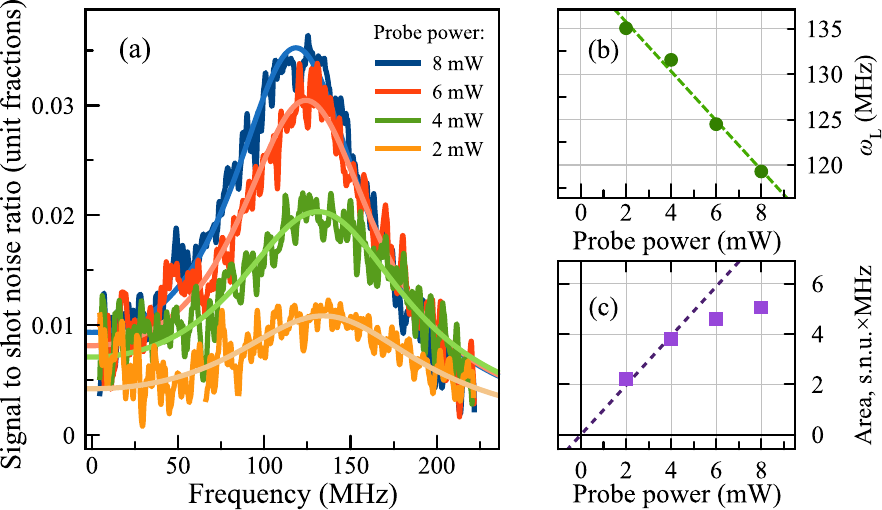}
\caption{(a) Dependence of spin noise on the probe power for $\lambda = 769.2$ nm, B = 21 mT. (b) Dependence of the Larmor frequency on the probe power, experimental data are shown with circles. The dashed line represents results of the calculation using the model described in Discussion. (c) Dependence of The peak area on the probe power. {The dotted line corresponds to dependence of the peak area for the case of purely unperturbing probe.}}
\label{sn-vs-power} 
\end{figure}

 To investigate the role of the system perturbation by the probe beam, we measured the spin noise spectrum at  fixed values of the wavelength and magnetic field ($\lambda =$~769.2~nm (1.612~eV), $B=21$~mT) for  different probe powers. The result is shown in Fig.~\ref{sn-vs-power}.  It demonstrates that {(i) the width does not change significantly, (ii) the position of the peak undergoes a slight linear shift and (iii) the area of the spectrum tends to saturate at higher power, while at low power (below 4~mW) it varies linearly}. Basically, this evidences the weakly perturbing nature of the SNS (the details are discussed below).

{Figs.~\ref{signal-vs-field} and~\ref{sn-vs-power} are obtained in the regime of the Faraday rotation noise detection.
However, we have checked that, for all wavelengths, temperatures, probe powers and polarization directions, these spectra (including their amplitudes) coincide with those of the ellipticity noise (SI, Fig.~6).
This behavior is not common for SNS since, for isotropic systems, the spectral dependencies of the ellipticity and Faraday rotation noise are governed by the absorption and refraction coefficients, respectively. In this case, their equality} is related to the known intermixing of the Faraday rotation and ellipticity signals in the birefringent crystals~\cite{kozlov-sn-birefringent22}.

\paragraph{Dependence of spin noise on magnetic field direction.}
	
\begin{figure*}
\centering
\includegraphics[width=\linewidth]{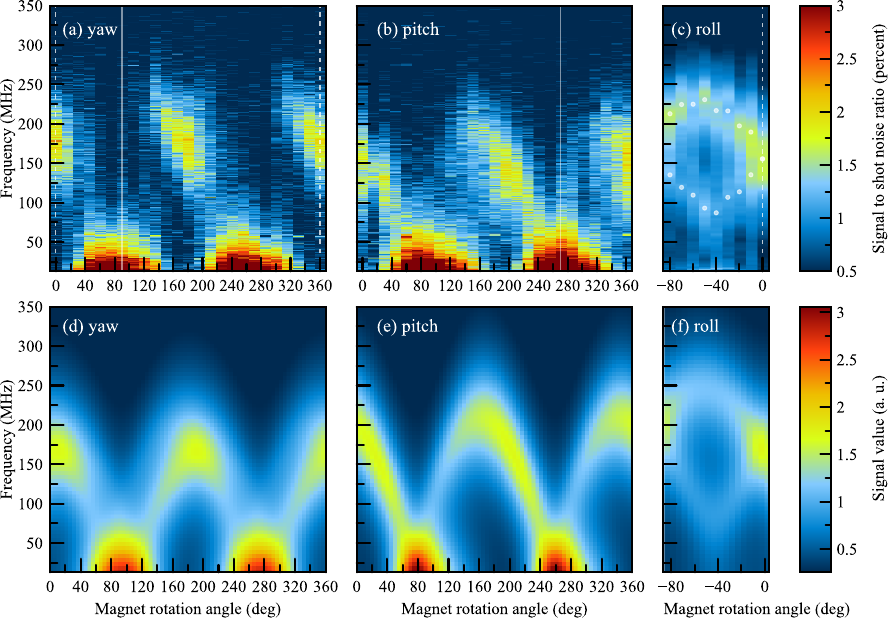}
\caption{Panoramic spin noise spectra in the rotating magnetic field of a fixed magnitude \textit{B}~$\approx$~20~mT, $\lambda$ = 769.5 nm, 9 mW of incident and 2 mW of transmitted light (see the SI Fig.~2a for definitions of pitch, yaw, and roll). (a) ``Yaw'' type rotation. (b) ``Pitch'' type rotation. On (a) and (b) panels, Voight (Faraday) geometry corresponds to $0 (90)\pm180^{\circ}$. (c) Rotation in the plane perpendicular to the optical axis (``roll''). Solid and dashed lines denote the coincident directions of magnetic field. Dots are extracted Larmor peak positions. (d-f) Calculations after model, presented in SI. See also the Figs.~7 in SI for more details.}
\label{field-angle} 
\end{figure*}

The MAPbI$_3$ crystal anisotropy leads to the anisotropy of the $g$\nobreakdash-factor~\cite{kirstein-spin-dyn-e-h-interact-nuclei-mapi-single-crystal22,garcia-arellano-unexp-anisotropy-g-factor-perovskite22}. To {reconstruct} the $g$\nobreakdash-tensor, we measured the spin noise spectra with 3D angular resolution of magnetic field direction. To do so, we performed three series of measurements while rotating the magnetic field of fixed magnitude around all three coordinate axes~\cite{kamenskii-invariants21}. {The crystal was not preliminary oriented, but the incident light was perpendicular to its (100) plane. To be concise, we will denote the rotations of magnetic field in the planes containing optical axis as the yaw and pitch, and the rotation in the plane normal to the optical axis as the roll. Additional information about the setup and sample geometry can be found in SI, Fig.~2a.} The configuration of our setup allowed us to make a full circle for the pitch and yaw, and to rotate the field by 80$^\circ$ for the roll ({in this case experiment always remains in the Voigt geometry}). The results are plotted as a color map in Fig.~\ref{field-angle}.

When the field was rotated between Faraday and Voigt geometry {by pitch or yaw}, the spectrum consisted of two peaks  centered at zero and Larmor frequencies, with their widths $\Delta\omega_0$ and $\Delta\omega_L$ corresponding to the longitudinal and transverse spin relaxation retes, $T_1^{-1}$ and $T_2^{-1}$, respectively ($T_1 \approx (\pi\Delta\omega_0)^{-1}$ and $T_2 \approx (\pi\Delta\omega_L)^{-1}$~\cite{smirnov-sn-review2021}). The longest relaxation times  $T_1$ and $T_2$, in our measurements, reached  $\sim \hskip1mm$4~ns.
The ratio of the extreme Larmor frequencies evaluated from Fig.~\ref{field-angle} yields the $g$\nobreakdash-factor anisotropy equal to 2.6, that well correlateswith the value estimated from the pump-probe Faraday rotation measurements~\cite{kirstein-spin-dyn-e-h-interact-nuclei-mapi-single-crystal22}.
        
Surprisingly, when the magnetic field was rotated in the Voigt geometry (by ``roll''), we observed splitting of the peak in the spin noise spectrum, which was the largest at the rotation angle of $\sim 45^\circ$. The ratios of the amplitudes and  frequencies of the two peaks did not depend on the probe power, polarization direction and temperature {(see also  Fig.~7 of SI)}. This evidences the two subensembles of the resident charge carriers with different principal axes of the $g$\nobreakdash-tensors.

\section{Discussion}
\paragraph{{The main information on spin subsystem retrieved by the SNS}.}
The spin noise spectroscopy, for the first time, was applied to investigate spin properties of an anisotropic semiconductor crystal, and for the first time to a {halide} perovskite crystal. Most of previous spin studies were performed in reflection geometry and did not reveal the volume-related phenomena like crystal twinning. The signal amplitude of several percent above shot noise is typical for the method and allows one to easily achieve panoramic spectra with varying magnetic field magnitude or direction, temperature, probe polarization and power and other  experimental parameters.

The measured values of the $g$\nobreakdash-factors point to the resident hole in the sample~\cite{kirstein-lande-factors-e-h-lead-halide-perovskites22,kirstein-spin-dyn-e-h-interact-nuclei-mapi-single-crystal22}.
{The observation of the holes at such low temperatures indicates the presence of shallow acceptor states in the sample.} Moreover, the spin noise spectra exhibit the features typical for the free charge carriers: (i) the shape of the peak is Lorentzian, so the spread of $g$\nobreakdash-factors is negligible, (ii) there is no zero-frequency peak in the Voight geometry, which is expected for the localized charge carriers at low magnetic fields~\cite{smirnov-sn-review2021}, (iii) $T_2$ time decreases with rising temperatures, which contradicts the hyperfine interaction driven spin relaxation reported for the localized charge carriers in halide perovskites~\cite{belykh-coherent-spin-dynamics-e-h-perovskite19,adma_Dortmund}. As a result, we conclude that the spin noise is produced by the resident delocalized holes.
This is in contrast with some  of previous investigations of the coherent spin dynamics in perovskites~\cite{adma_Dortmund,garcia-arellano-energy-tuning-e-spin-perovskites21,kirstein-spin-dyn-e-h-interact-nuclei-mapi-single-crystal22} and can be related to the fact that our measurements were performed in  the transmission (rather than reflection) geometry.
Moreover, the observed spin coherence time of $T_2=4$~ns is the longest reported in perovskites so far. We assume that both delocalization of the charge carriers and high crystal quality are essential for prolonging spin relaxation times. The latter is achieved in our case by counterdiffusion-in-gel growth method instead of the growth from  solution.

\paragraph{{Limits of the non-perturbative measurements.}}
The increase of the probe beam intensity weakly perturbs the resident hole spin dynamics, as shown in Fig.~\ref{sn-vs-power}. To explain this, we note that the selection rules for the optical transitions in perovskite crystals are similar to the optical transitions from $\Gamma_7$ to $\Gamma_6$ band in GaAs-like semiconductors~\cite{odenthal-spin-pol-q-beating-perovskites2017}. They determine virtual interband transitions, which produce the resonant spin noise signal. In Supplementary Information, we present a general framework for theoretical description of the optical spin noise spectroscopy. However, we find that the experimental results can be described by a particular limiting case of strong exchange interaction between resident and photoexcited hole, which dominates over the interaction between the photoexcited hole and electron.

In this limit, absorption of a (virtual) photon leads to creation of a singlet trion, which consists of two holes with antiparallel spins and an electron with unpaired spin. Hamiltonian of the system in a magnetic field parallel to the $x$ axis reads:
\begin{equation}
\mathcal H=\hbar\Omega_hS_x+\hbar\omega_0n_{tr}+\hbar\Omega_e S_{e,x}+\left(d_x\mathcal E\ee^{-\i\omega t}+{\rm{H.c.}}\right).
\end{equation}
Here, $\bm S$ is the spin of the resident hole, $\hbar\Omega_h$ is the hole Zeeman splitting, $\omega_0$ is the trion resonance frequency, $n_{tr}$ is the occupancy of the trion state, $\bm S_e$ is the spin of electron in the trion, $\hbar\Omega_e$ is its Zeeman splitting, $d_x$ is the trion optical transition dipole moment operator corresponding to the light polarized along the $x$ axis, $\mathcal E$ is proportional to amplitude of the probe light, and $\omega$ is the probe light frequency. The steady state occupancy of the trion state is of the order of $(\mathcal E\gamma_0)^2/(\omega_0-\omega)^4$, where $\gamma_0$ is the homogeneous width of the trion resonance. Assuming it to be small, we find that the spin noise spectrum a standard Lorentzian shape, though with a renormalized resonance frequency
\begin{equation}
\Omega=\Omega_h+\frac{\mathcal E^2(\Omega_e-\Omega_h)}{2(\omega_0-\omega)^2}.
\end{equation}
This expression describes the renormalization of the precession frequency due to the difference of the $g$\nobreakdash-factors of a trion and a resident hole. This happens because the fast trion creation and recombination lead to a single common peak in the spin noise spectrum. In the MAPbI$_3$ crystal, the electron has a larger $g$\nobreakdash-factor than the hole and of  different sign, so $\Omega_e\approx-9\Omega_h$~\cite{kirstein-lande-factors-e-h-lead-halide-perovskites22}. Thus, the increase of the probe power leads to  linear decrease of the precession frequency. The corresponding fit is shown in Fig.~\ref{sn-vs-power} in the top right corner, which yields $\mathcal E^2/(\omega_0-\omega)^2=P/$(260~mW), with $P$ being the probe power.

The slow hole spin relaxation is likely limited by the Dyakonov-Perel spin relaxation mechanism~\cite{dyakonov72,dyakonov_book}. The relaxation rate is given by $T_2^{-1}=\left<\Omega_{\rm s.o.}^2\right>\tau_p$, where $\left<\Omega_{\rm s.o.}^2\right>$ is the average hole spin precession frequency due to the spin-orbit interaction squared, and $\tau_p$ is the momentum relaxation time. For the non-degenerate hole gas, the spin relaxation time can be estimated as
\begin{equation}
T_2\sim\frac{\hbar^4 e}{\Lambda^2\mu m^2 k_B T},
\end{equation}
where $\Lambda$ is the spin-orbit coupling constant, $\mu$ is the mobility, and $m$ is the hole mass. For the realistic parameters $m_h=0.1m_e$ with $m_e$ being the free electron mass, $\mu=0.1$~cm$^2$/(V$\cdot$s), $\Lambda=1$~eV$\cdot$\AA~\cite{doi:10.1073/pnas.1405780111,doi:10.1021/acsnano.5b04409}, and $T=3$~K we obtain $T_2\sim$10~ns in {relatively good} agreement with the experimental value.

\paragraph{The origin of spin subensembles.}
The anisotropy of the $g$\nobreakdash-factor is related to the anisotropy of the crystal at low temperatures. So we suggest that the two subensembles originate from the twinning of the crystal. This suggestion is supported by the previous observations of the domains in MAPbI$_3$ crystal by Laue neutron diffraction~\cite{breternitz-twinning-mapbi3-room-temp20}, TEM and SEM imaging~\cite{rothmann-direct-observation-intrinsic-twin-domain-mapi17} and ferroelastic-based measurements~\cite{strelcov-mapi-perovskites-ferroelasticity17,liu-twin-domains-mod-light-matter-inter-halide-perovskites20}. The size of the domains and their distribution can be further investigated using different growth conditions. {However, we 
performed the modelling of the panoramic spin noise spectra (Fig.~\ref{field-angle}d-f) and determined, as far as possible, the directions of the axes in the domains formed as a result of twinning. We started from the following assumptions.}
\begin{enumerate}
\item {In Voight geometry only the Larmor frequency signal is observed, and in the Faraday geometry resides only the zero frequency peak. It means that the anisotropy axes lie in a plane (almost) perpendicular to the direction of light propagation~\cite{kamenskii-invariants21}.}
\item {With the ``roll'' rotation, we observe the frequency variations of Larmor peaks, which are almost in the opposite phase. It means that the angle between anisotropy axes is close to $\frac{\pi}{2}$.}
\item {The maximum splitting of the peaks occurs close to --45$^\circ$ angle of the ``roll'' rotation, indicating that the anisotropy axes are inclined at 45$^\circ$ to the polarization of the input light, which is vertical with respect to the surface of the optical table.}
\end{enumerate}
{With such assumptions, the resulting modeling plots become highly symmetrical, which required to slightly modify the parameters to better match the experimentally observed picture. The parameters used are given in the SI.} {They correspond to the longitudinal and transverse $g$-factors in each domain equal to 0.9 and 0.3, respectively.}

\section{Conclusions}

{In this work, thanks to the capabilities of the SNS for working with birefringent media, we} obtained for the first time the spin noise spectrum of a perovskite MAPbI$_3$ single crystal at low temperature in the orthorhombic phase. We observed an equilibrium spin noise of resident free holes, and measured a record long spin dephasing time of $T_2=4$~ns. {With rising probe power, when SNS deviates from it nonperturbative nature, we} found the effect of the precession frequency renormalization due to the virtual excitation of the electron-hole pairs, which allowed us to build an effective Hamiltonian of the ground and excited states. We also performed the SNS with angular resolution of magnetic field direction, which gave access to the anisotropy of the resident holes $g$\nobreakdash-factor and longitudinal spin relaxation time. {Finally, w}e observed splitting of the spin noise spectrum for a certain direction of magnetic field, which reveals spontaneous twinning of the crystal.

Results of this work not only provide new data about properties of a particular spin-system, but also demonstrate a number of unusual abilities of this experimental technique. Among them are applicability of the technique to strongly birefringent crystals, specific measurements with external-field manipulations, high spatial resolution of the spin-resonance measurements, detecting spin resonance under conditions of optical pumping. {For the first time, the twinning of a bulk crystal was observed by means of spin noise spectroscopy. The results of our work advance the research field of these promising semiconductor systems, opening up a way of non-perturbative monitoring of its spin subsystem state and dynamics and investigation of spin-related phenomena with optical spatial and spectral resolution.}

\section*{Acknowledgments}
	
The main experimental work (sample synthesis and spin noise spectroscopy measurements) was financially supported by the Ministry of Science and Higher Education of the Russian Federation Megagrant №075-15-2022-1112, which is highly appreciated. The sample birefringency measurements were done under the support of the Saint Petersburg State University (Grant No. 94030557). The theoretical description of the spin noise spectra by D.S.S. were supported by the Russian Science Foundation Grant No. 23-12-00142. The sample growth and experimental work was fulfilled using the equipment of Resource Center ``Nanophotonics'' of Saint Petersburg State University Research Park.

\onecolumngrid
\newpage

\appendix

\setcounter{figure}{0}

\section*{Methods}

\subsection*{Samples under study}
The samples under study were bulk MAPbI$_3$ crystals grown by the counterdiffusion-in-gel method~\cite{selivanov-counterdiffusion-in-gel-growth-perovskite22}. These crystals demonstrate relatively better optical and spin properties in terms of linewidths and lifetimes. The crystals from the same growth run were preliminary studied using the Faraday rotation (FR) method~\cite{shumitskaya-fr-phase-trans-perovskite-2023-preprint}. Temperature dependence of the detected FR signal unambiguously indicated its paramagnetic nature and, therefore, the presence of spin carriers and potential possibility of SNS measurements. 

The MAPbI$_3$ samples are difficult to polish; therefore, in the experiments, the crystals have a rather inhomogeneous surface and the transmitted light becomes mainly scattered. The better optical quality in some cases can be achieved by chipping the crystal. The possibility of spin noise observation depends very strongly on the probed point on the sample.

An important circumstance of the studies was the fact that after one cycle of cooling the sample to the operating temperature and subsequent heating, degradation of the samples occurred, which was expressed in an increase in the fraction of scattered light and the disappearance of the spin noise signal upon repeated cooling. In this case, there was no visible destruction of the surface, however, when the crystals were removed from the holder, they began to crumble. In our setup, the crystals were attached to a cold copper finger using silver paste, both of these materials are aggressive to the MAPbI$_3$ crystal, but provide the best thermal contact. Another possible reason for the destruction of the sample may be prolonged vacuum pumping, which can lead to local evaporation of methylammonium and disruption of the structure.

The paper presents experimental data from one MAPbI$_3$ crystal, which were completely obtained in one cooling cycle. Their reproducibility was further confirmed by verification measurements of several other crystal fragments from the same growth series.

\subsection*{Optical characterization}
\begin{figure}[h]
 \centering
 \includegraphics[width=\linewidth]{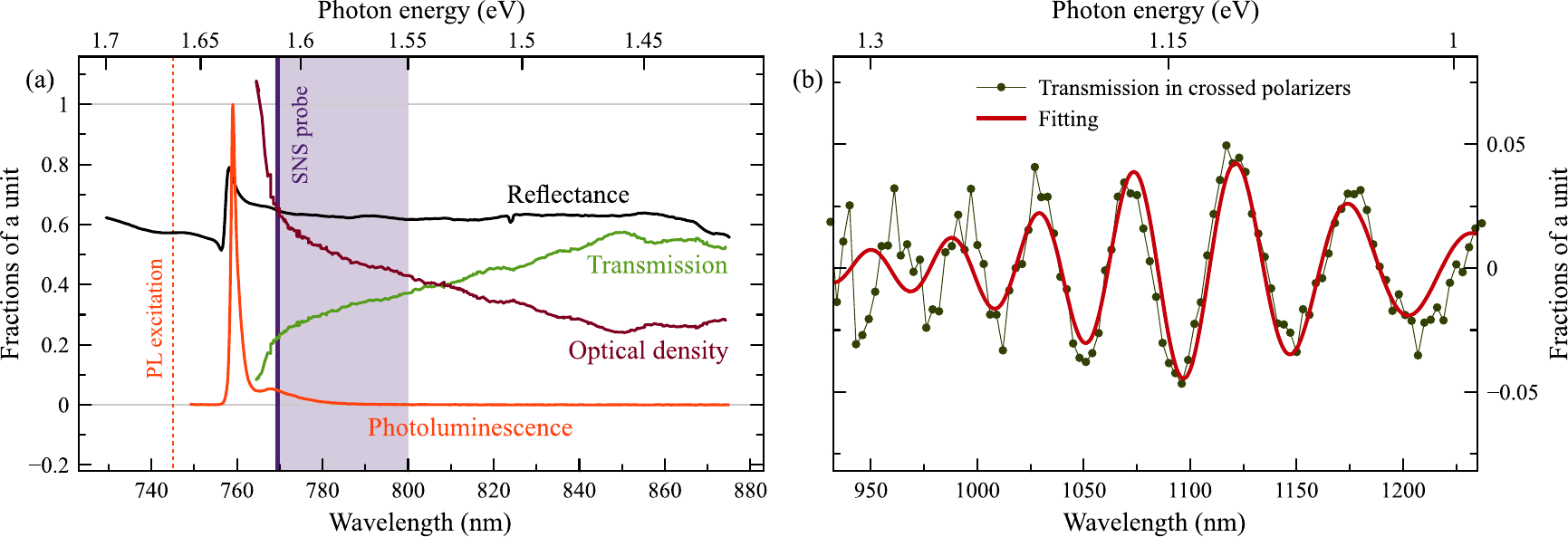}
 \caption{\label{optics} (a)~Transmission, optical density, photoluminescence (PL) and reflectance spectra of the samples under study. The tinted area indicates the wavelength range at which the spin noise was measured (thick vertical line indicates the main wavelength in most experiments). The PL spectrum is notmalized to maximum value, excitation wavelength is 745~nm (1.66~eV). (b)~The transmission spectrum of a sample placed between crossed polarizers at room temperature (backgorund is subtracted).
 }
\end{figure}

\begin{figure}
 \centering
 \includegraphics[width=\linewidth]{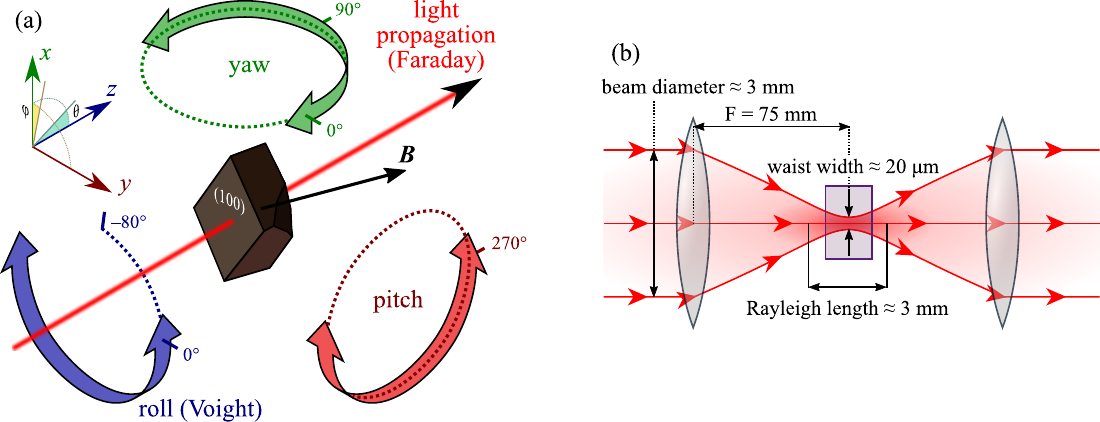}
 \caption{(a)~The SNS setup geometry and notation of the rotation axes. (b)~Approximate parameters of the waist width and Rayleigh length on the sample are given. Assuming that a light beam with a diameter (D) of 3 mm, of uniform intensity, is incident on the lens, and the waist width is determined by the diffraction limit. The Rayleigh length is estimated for a Gaussian beam.}
 \label{geometry}
\end{figure}

The transmission (and the corresponding optical density) spectrum of the crystal (Fig.~\ref{optics}a) was measured at the point on the sample, where the spin noise was then recorded. The spectrum was obtained by tuning the laser wavelength and simultaneous measurement of the incident and transmitted light powers. The optical range is determined by the range of the mirrors of the TIS-SF-777 titanium-sapphire laser used. We also present photoluminescence and reflectance spectra obtained from other samples of the same growth series. All the spectra of Fig.~\ref{geometry}a were collected at 4~K.

The MAPbI$_3$ crystal is birefringent at temperatures below 327~K ($\sim 54^\circ$C), and the light, initially linearly polarized, transmitted through the sample generally appears to be elliptically polarized. Here, we made an attempt To evaluate the sample birefringence. For this purpose, we measured transmission spectrum of the sample placed between two crossed polarizers 
in the Cary spectrometer (Fig.~\ref{optics}b). Under these conditions, the transmitted light proves to be modulated in accordance with the equality $\Delta k l \Delta n = 2\pi$, where $\Delta k = \frac{2\pi}{\Delta\lambda}$, $\Delta\lambda$ is the wavelength difference between two adjacent transmission extrema, $l$ is the sample thickness, $\Delta n$ is the refractive index difference for the two normal waves. It is noteworthy that such factors as smallness of the sample, its low optical quality. and inaccuracy in optical alignment of the sample could significantly reduce amplitude of the spectral modulation, but could not affect  spectral period of its modulation, which was of paramount importance for these measurements.  The obtained spectrum was fitted by a sinusoidal function of the inverse wavelength with an arbitrary envelope. $\Delta n = \frac{1}{\Delta\xi l}$, where $\xi = \lambda^{-1} = 4.12\cdot10^{-5}$nm$^{-1}$, $l \approx 1.3\cdot10^{6}$nm. Thus, in the MAPbI$_3$ crystal, $\Delta n \approx 0.0185$ at room temperature. The approximation of Fig.~\ref{optics}b also shows a slight variation of the modulation period across the spectrum that can be used to evaluate dependence of the birefringence $\Delta n$ on the wavelength.

The SNS setup geometry is shown in Fig.~\ref{geometry}a. the light propagates along $z$ axis. We denote the rotation of magnetic field around $z$ axis ($xy$ plane) as ``pitch'' (in this case any angle corresponds to Voight geometry), rotation in $xz$ plane as ``pitch'' and rotation in $zy$ plane as ``yaw''. The 0$^{\circ}$ angles for ``yaw'' and ``roll'' coincide, and the 270$^{\circ}$ ``pitch'' angle technically corresponds to 90$^{\circ}$ ``yaw'' angle. The $\theta$ ($\phi$) denotes the anisotropy axis angle in the $xz$ plane, measured from $z$ ($x$) axis, and $\phi$ denotes the anisotropy axis angle in the $xy$ plane, measured from $x$ axis (this notation is used in model of the field angle dependence). The Fig.~\ref{geometry}b presents the approximate geometrical properties of the focused beam.

\subsection*{Spin noise measurements}
The spin noise signal was obtained at liquid helium temperatures using the common SNS setup (see,~e.~g.,~\cite{glazov-linear-optics-raman-sns2015}). The light source was a tunable ring cavity CW Ti:sapphire laser  TIS-SF-777. The samples were placed in a closed-cycle cryostat Montana Cryostation with an electromagnet or, alternatively, a permanent rotating magnet outside the chamber (for magnetic field orientation measurenents). The polarimetric setup included the broadband balanced photoreceiver Newport NewFocus 1607-AC. The signal was collected with use of fast Fourier transform (FFT) based radiofrequency (RF) spectrum analyzer Tektronix RSA5103A. To record ellipticity noise spectra, a $\lambda/4$ plate was placed after the sample. One of the optical axes of quarter waveplate was aligned along the main axis of elliptical polarization of the light transmitted through the sample.

\newpage

\section*{Supplementary information}

Here, we present additional spin noise measurement data in the experimental section, and the developed model in the theoretical section.

\subsection*{Experimental}

\begin{figure}[h]
	\centering
	\includegraphics[width=0.9\linewidth]{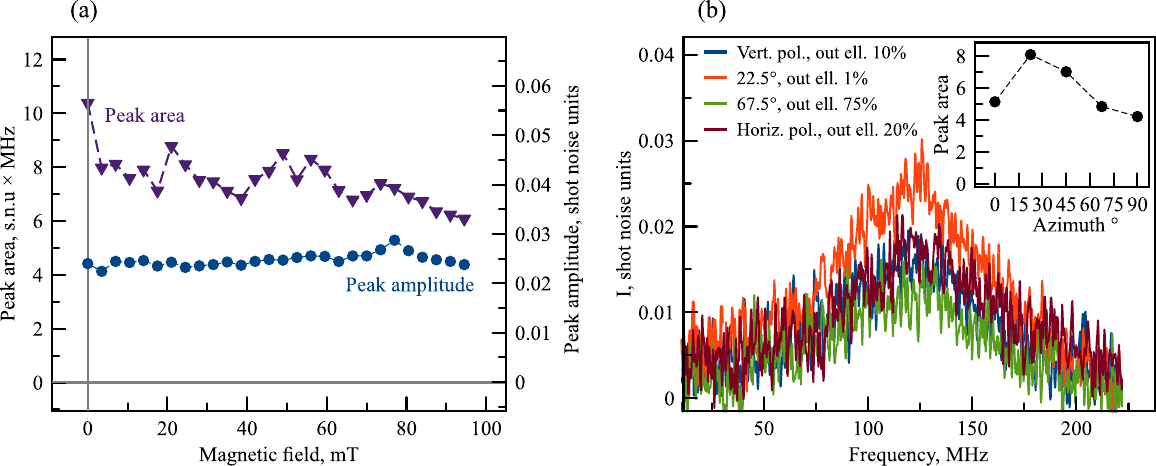}
	\caption{(a)~The area and amplitude of the spin noise Larmor peak as a function of the magnitude of the applied magnetic field. (b)~Spin noise spectrum dependence on polarization azimuth. Probe light power 8 mW, $\lambda = 769.2$ nm, B = 21 mT.
	}
	\label{field-and-pol} 
\end{figure}

\paragraph{Spin noise dependence on magnetic field}
In a nonzero magnetic field, the area of the spin noise signals remains virtually unchanged (Fig.~\ref{field-and-pol}a) within the spread. 

\paragraph{Dependence of spin noise on the azimuth of the plane of polarization}
The change of incident polarization plane azimuth does not lead to any transformation of the signal, except for a decrease of its amplitude, while the output polarization becomes elliptical (Fig.~\ref{field-and-pol}b).
The inset shows the dependence of the peak area on the polarization azimuth. The polarization was changed by rotating the polarizer axis in front of the sample while maintaining the power of the light incident on the sample at 8 mW. The amplitude of the observed signal changed only in proportion to the change in the ellipticity of the light emerging from the birefringent sample and the corresponding decrease in the sensitivity of the detector to the rotation signal.

\begin{figure}
	\centering
	\includegraphics[width=0.9\linewidth]{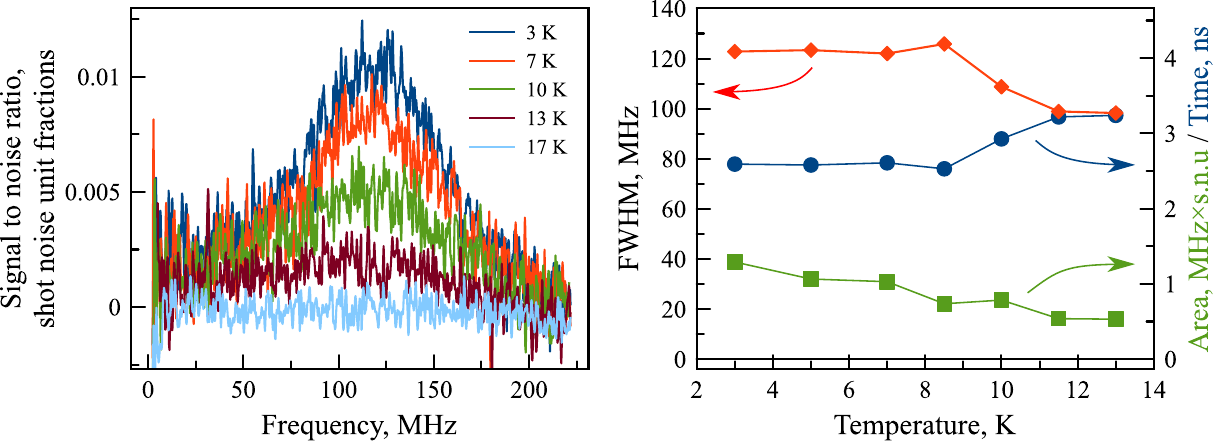}
	\caption{(a)~Temperature dependence of spin noise. Probe light power P$_{\mbox \small in}$= 8 mW, $\lambda = 769.2$ nm, B = 21 mT. (b)~Normalized to light power area of spin noise peak per square of output power as a function of temperature. The inset shows the normalized to light power spin noise amplitude.
	}
	\label{temperature} 
\end{figure}

\paragraph{Temperature dependence of spin noise}
The temperature dependence of the spin noise spectrum is shown in Fig.~\ref{temperature}a. The measurements were carried out in the range from 3 to 17 K; at higher temperatures, the signal became unobservable. Since the output power varied with temperature, Fig.~\ref{temperature}b shows the normalized to light power value of the spin noise peak area as a function of temperature.

\begin{figure}
 \centering
 \includegraphics[width=0.9\linewidth]{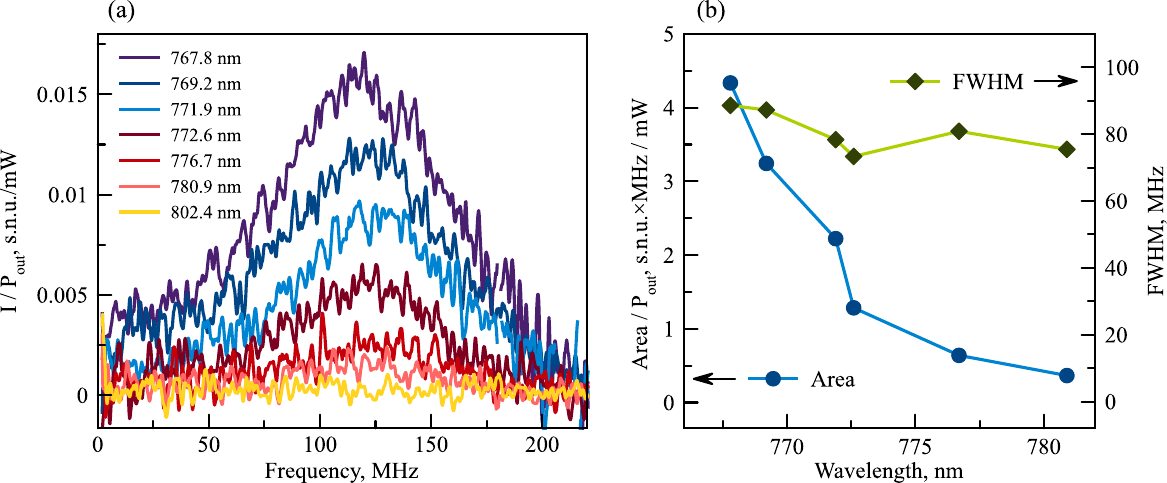}
 \caption{(a)~Normalized spectrum of spin noise as a function of wavelength. Input power 8 mW, B = 21 mT. (b)~Area and FWHM of spin noise (normalized to transmitted power) as a function of wavelength.} 
 \label{sn-vs-wl}
\end{figure}

\paragraph{Spin noise dependence on wavelength}
To get rid of the transmission specral variation, we normalized the spin noise to the square of the output power. Fig.~\ref{sn-vs-wl} shows the dependence of spin noise spectra (a), its area and width of the normalized signal (b)~on the wavelength.

\begin{figure}
	\centering
	\includegraphics[width=0.9\linewidth]{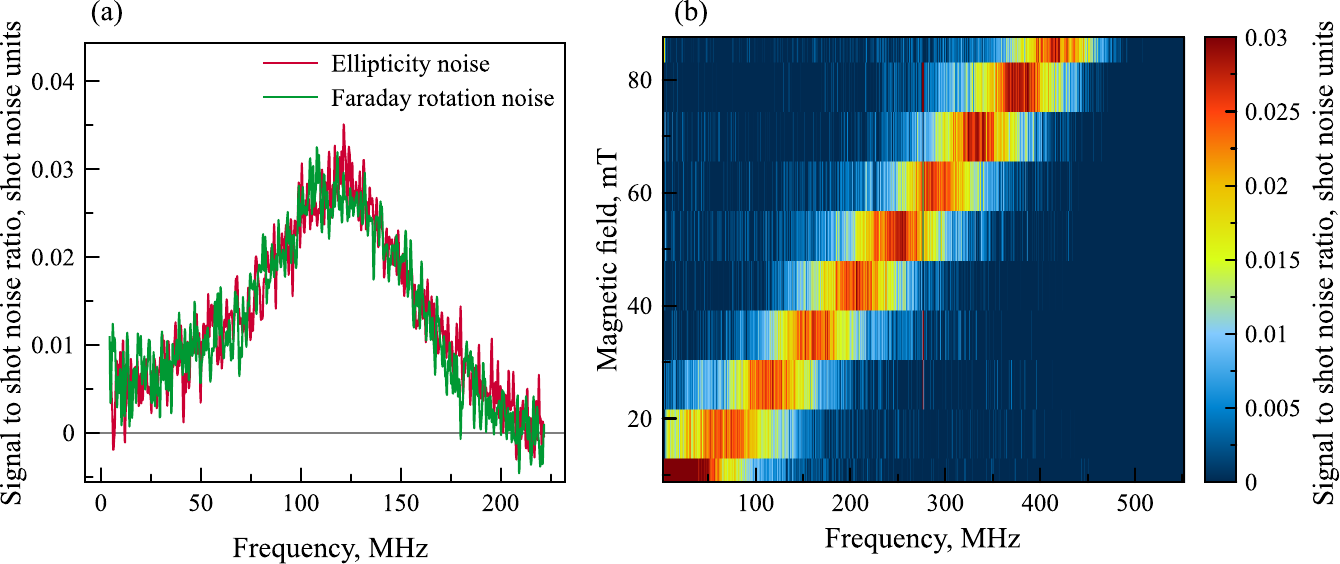}
	\caption{(a)~Ellipticity noise and Faraday rotation noise. Input power P$_{\mbox \small in}$ 8 mW, $\lambda = 769.2$, nm, B = 21 mT. (b)~Dependence of the ellipticity noise on the magnetic field. Input power P$_{\mbox \small in}$ = 8 mW, $\lambda = 769.2$.
	}
	\label{ellipticity} 
\end{figure}

\paragraph{Spin noise spectrum in ellipticity noise}
Ellipticity noise (EN) and Faraday rotation noise (FRN) spectra were found to be completely identical in all experimental conditions (Fig.~\ref{ellipticity}a). EN and FRN signals demonstrated the same behavior for any set of controlled parameters, for example, when changing the magnitude of the external magnetic field, we observe the same pattern in the ellipticity noise (Fig.~\ref{ellipticity}b) as in the noise of the Faraday rotation (see Fig.~1 in the main text). This is in good agreement with previous spin noise measurements in birefringent media~\cite{kozlov-sn-birefringent22}.

\noindent

\begin{figure}
	\centering
	\includegraphics[width=\linewidth]{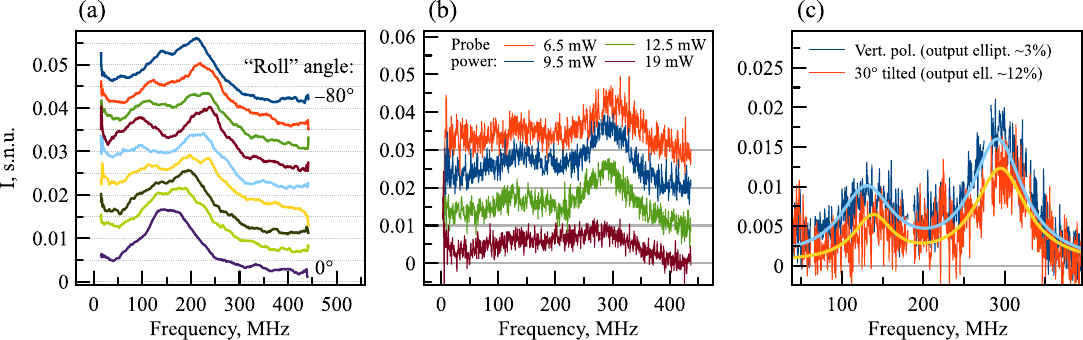}
	\caption{(a)~Spin noise spectra dependence on the direction of the magnetic field. Rotation in the plane perpendicular to the optical axis (``roll''). Probe light power 9 mW, $\lambda = 769.5$ nm, B $\approx$ 20~mT. Graphs are shifted by 0.05 along y-axis. (b)~Spin noise spectrum depending on the power of light probing the medium. Probing light $\lambda = 770.2$ nm, B $\approx$ 23 mT, magnetic field direction angle $= -45^\circ$. Graphs are shifted by 0.1 along y-axis. (c)~Spin noise spectrum at two different azimuth angles of polarization of light probing the medium. Probe light power 9.5 mW, $\lambda = 769.5$ nm, B $\approx$ 23 mT, magnetic field direction angle $= -45\deg$.
	}
	\label{splitting} 
\end{figure}

\paragraph{Spin noise twinning analysis}
Fig.~\ref{splitting}a shows {the data from Fig.~3(c)~of the main paper, plotted as waterfall for better perception.}
When the probe power changes, the ratio of the areas between the split peaks remains constant (Fig.~\ref{splitting}b), a similar behavior is revealed when the polarization azimuth angle of the incident light changes (Fig.~\ref{splitting}c).

\paragraph{Modelling the spin noise intensity with account for the twinning}

The spin noise intensity in Fig. 3(d--e) in the main text was calculated as a sum over the two domains: $(\mathcal F^2)_\omega=(\mathcal F^2)_\omega^{(1)}+(\mathcal F^2)_\omega^{(2)}$, where~\cite{smirnov-review21en}
\begin{equation}
    (\mathcal F^2)_\omega^{(i)}=A_i \left\{ \cos^2\alpha_i \Delta_{T_1}(\omega) +\frac{1}{2}\sin^2\alpha_i\left[\Delta_{T_2}(\omega-\Omega_i)+\Delta_{T_2}(\omega+\Omega_i)\right]\right\}
\end{equation}
with $i=1,2$. Here $A_i$ is an amplitude of the corresponding contribution, $\bm\Omega_i=\hat{g}_i\mu_B\bm B/\hbar$ with $\hat{g}_i$ being the tensor of $g$-factors, $\mu_B$ being the Bohr magneton, and $\bm B$ being external magnetic field, $\alpha_i$ is an angle between $\bm\Omega_i$ and the light propagation direction, $T_1$ and $T_2$ are the longitudinal and transverse spin relaxation times, respectively, and
\begin{equation}
    \Delta_T(\omega)=\frac{1}{\pi}\frac{T}{1+(\omega T)^2}.
\end{equation}
We assume the $g$-tensors to be uniaxial with the directions of the main axes described by the angles $\theta_i$ and $\varphi_i$, see the inset in Fig.~\ref{geometry}, component $g_\parallel$ along the main axis and components $g_\perp$ in the perpendicular directions.

For Fig. 3(d--e) in the main text we used the following parameters: $g_\parallel=0.9$, $g_\perp=0.3$, $T_1=T_2=2.6$~ns, $A_1/A_2=1.7$, the directions of the main axes in the two domains are determined by the angles $\theta_1=\pi/2-\pi/12$, $\phi_1=-3\pi/4-\pi/18$, $\theta_2=\pi/2+\pi/12$, and $\phi_2=-\pi/4$ in spherical coordinates, see the inset in Fig.~\ref{geometry}(a).

\newpage

\subsection*{Theory}

\newcommand{\Tr}{\mathop{\rm Tr}\nolimits}
\renewcommand{\Re}{\mathop{\rm Re}}
\renewcommand{\i}{{\rm i}}
\renewcommand{\d}{\mathrm d}
\newcommand{\bbraket}[1]{\left\langle #1 \right\rangle}

\subsubsection*{General model}

We consider a single hole in a perovskite crystal in external transverse magnetic field (Voight geometry). The coherent optical excitation by linearly polarized light leads to the real or virtual creation of an additional electron-hole pair (exciton), which interacts with the resident hole. The Hamiltonian of the system can be written as ($\hbar=1$)
\begin{equation}
  \mathcal H=\Omega_hS_x+\omega_0n_{ex}+\left(d_x\mathcal E\ee^{-\i\omega t}+{\rm{H.c.}}\right)+J\bm S\bm S_h'+\Delta\bm S_e'\bm S_h'+\Omega_h S_{h,x}'+\Omega_e S_{e,x}',
\end{equation}
where $\Omega_{h,e}$ are the hole and electron spin precession frequencies in the external magnetic field, $\bm S$ is the resident hole spin, $\bm S_h'$ and $\bm S_e'$ are the hole and electron spins in the exciton, $\omega_0$ is the exciton optical transition frequency, $n_{ex}$ is the occupancy of the exciton state, $d_x$ is proportional the optical transition dipole moment component along $x$ axis, $\mathcal E$ is proportional to the amplitude of the incident field, $\omega$ is its frequency, $J$ is the exchange interaction constant between resident and photoexcited hole, $\Delta$ is the strength of the electron-hole exchange interaction in the exciton. The incoherent processes can be described using the density matrix $\rho$, which satisfies the Lindblad equation
\begin{equation}
  \label{eq:Lindblad}
  \frac{\d\rho}{\d t}=-\i\left[\mathcal H,\rho\right]-\mathcal L\left\{\rho\right\},
\end{equation}
where
\begin{equation}
  \mathcal L\left\{\rho\right\}=\gamma\sum_{i=\pm,z}\left(d_i^\dag d_i\rho+\rho d_i^\dag d_i-2d_i\rho d_i^\dag\right)+\frac{1}{2\tau_s}\sum_{i=x,y,z}\left(\rho/2-2S_i\rho S_i\right).
\end{equation}
Here $\gamma$ is the radiative recombination rate and $\tau_s$ is the hole spin relaxation time in the ground state. The operators, which describe the dipole moments $d_+$, $d_-$, and $d_z$ are the annihilation operators of the following states (in the basis $\left|S_{h,z}',S_{e,z}\right>$):
\begin{equation}
  \left|+\frac{1}{2},+\frac{1}{2}\right>,
  \qquad
  \left|-\frac{1}{2},-\frac{1}{2}\right>,
  \qquad
  \frac{1}{\sqrt{2}}\left(\left|+\frac{1}{2},-\frac{1}{2}\right>+\left|-\frac{1}{2},+\frac{1}{2}\right>\right).
\end{equation}
Moreover, $d_x=(d_++d_-)/\sqrt{2}$. Additionally, the nonradiative recombination and the spin relaxation of the photoexcited charge carriers can be taken into account.

The Faraday rotation signal has the form
\begin{equation}
  \mathcal F\propto|E_{x'}^2|-|E_{y'}^2|\approx2\Re(E_yE_x^*)
\end{equation}
where $\bm E$ is the electric field of the transmitted light and $E_{x'}=(E_x+E_y)/\sqrt{2}$, $E_{y'}=(E_y-E_x)/\sqrt{2}$. To calculate the correlation function of the Faraday rotation signals one has to consider the normal and time ordering of the operators of the electric field $\hat{\bm E}$:
\begin{equation}
  \bbraket{\mathcal F(0)\mathcal F(t)}\propto2\Re\bbraket{\hat E_x^\dag(0)\hat E_x^\dag(t)\hat E_y(t)\hat E_y(0)+\hat E_y^\dag(0)\hat E_x^\dag(t)\hat E_y(t)\hat E_x(0)}.
\end{equation}
The electric field component along the $x$ axis can be replaced with the classical electric field $\mathcal E\ee^{-\i\omega t}$, so we obtain
\begin{equation}
  \bbraket{\mathcal F(0)\mathcal F(t)}\propto\Re\Tr\left\{d_y(t)\left[d_y(0)\rho\ee^{2\i\omega t}+d_y^\dag(0)\rho\right]\right\},
\end{equation}
where $d_y=\i(d_--d_+)/\sqrt{2}$. This average can be calculated in the Shr\"odinger representation from the solution of Eq.~\eqref{eq:Lindblad} with the initial condition $\rho(0)=d_y\rho_0$, where $\rho_0$ is the steady state density matrix, as follows:
\begin{equation}
  \label{eq:corr}
  \bbraket{\mathcal F(0)\mathcal F(t)}\propto\Re\Tr\left[\left(d_y^\dag+d_y\ee^{2\i\omega t}\right)\rho(t)\right].
\end{equation}

\subsubsection*{Description of the experimental results}

We find that the experimental spin noise spectra can be well described by a simplified model, which is similar to the spin noise in the quantum dots. Namely, we assume that the spin noise signal of holes in perovskite is dominated by the singlet trion resonance. This limit can be obtained from the general model by setting $J\gg\Delta$ and $\omega\approx\omega_0-(3/4)J$. In this limit, the spins of two holes in the excited state compensate each other, so the trion spin dynamics is determined by the unpaired electron spin. As a result, the Hamiltonian can be rewritten as
\begin{equation}
  \label{eq:Ham_trion}
  \mathcal H=\Omega_hS_x+\omega_0n_{ex}+\left(d_x\mathcal E\ee^{-\i\omega t}+{\rm{H.c.}}\right)+\Omega_e S_{e,x},
\end{equation}
where the hole spin operator $\bm S$ acts only in the ground state, and $\bm S_e$ acts only in the excited state. The Lindblad superoperator can be written in this limit as
\begin{equation}
  \mathcal L\left\{\rho\right\}=\frac{\gamma_0}{3}\sum_{i=\pm,z}\left(d_i^\dag d_i\rho+\rho d_i^\dag d_i-2d_i\rho d_i^\dag\right)+\frac{1}{2\tau_s}\sum_{i=x,y,z}\left(S_i^2\rho+\rho S_i^2-2S_i\rho S_i\right),
\end{equation}
and the operators proportional to the dipole moment have the following nonzero matrix elements:
\begin{multline}
  \left<S_z=-1/2\middle|d_+\middle|S_{e,z}=+1/2\right>=1,
  \qquad
  \left<S_z=+1/2\middle|d_-\middle|S_{e,z}=-1/2\right>=1,\\
  \left<S_z=+1/2\middle|d_z\middle|S_{e,z}=+1/2\right>=1/\sqrt{2},
  \qquad
  \left<S_z=-1/2\middle|d_z\middle|S_{e,z}=-1/2\right>=1/\sqrt{2}.
\end{multline}

\begin{figure}
  \centering
  \includegraphics[width=0.49\linewidth]{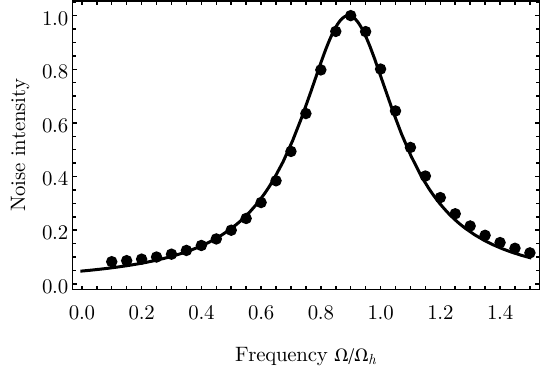}
  \caption{\label{fig}
    Comparison between numeric calculation after Eqs.~\eqref{eq:Ham_trion}, \eqref{eq:corr} and analytical expressions~\eqref{eq:F2}, \eqref{eq:Omega} for $\delta=750\Omega_h$, $\mathcal E=110\Omega_h$, $\gamma_0=0.05\Omega_h$, $1/\tau_s=0.2\Omega_h$, $\Omega_e=-9\Omega_h$.
  }
\end{figure}

We consider the limit of $\gamma_0,\mathcal E\ll|\delta|$, where $\delta=\omega_0-(3/4)J-\omega$ is the detuning from the trion resonance. In this limit, the occupancy of the trion state $\sim(\mathcal E\gamma_0/\delta^2)^2$ is very small. However, the optical excitation leads to the renormalization of the precession frequency (and smaller renormalization of the spin relaxation time) in the ground state, which gives the standard form of the spin noise spectrum
\begin{equation}
  \label{eq:F2}
  (\mathcal F^2)_\omega\propto\frac{\tau_s}{1+(\omega-\Omega)^2\tau_s^2},
\end{equation}
with the renormalized frequency
\begin{equation}
  \label{eq:Omega}
  \Omega=\Omega_h+\frac{\mathcal E^2}{2\delta^2}(\Omega_e-\Omega_h).
\end{equation}
The comparison between this analytical expression and numeric calculation is shown in Fig.~\ref{fig}.

The $g$ factor of electrons in MAPbI$_3$ is approximately 9 times larger than that of the holes and has an opposite sign, so $\Omega_e=-9\Omega_h$. This leads to the decrease of the spin precession frequency with increase of the excitation power:
\begin{equation}
  \label{eq:fit}
  \Omega=\Omega_h\left(1-5\frac{\mathcal E^2}{\delta^2}\right).
\end{equation}

\end{document}